# BUILDING SCENARIOS FOR ENVIRONMENTAL MANAGEMENT AND PLANNING: AN IT-BASED APPROACH[1]

by Dino Borri and Domenico Camarda[2]

Dipartimento di Architettura e Urbanistica, Politecnico di Bari, Italy.

**Abstract**

Oftentimes, the need to build multidiscipline knowledge bases, oriented to policy scenarios, entails the involvement of stakeholders in manifold domains, with a juxtaposition of different languages whose semantics can hardly allow inter-domain transfers.

A useful support for planning is the building up of durable IT-based interactive platforms, where it is possible to modify initial positions toward a semantic convergence.

The present paper shows an area-based application of these tools, for the integrated distance-management of different forms of knowledge expressed by selected stakeholders about environmental planning issues, in order to build alternative development scenarios.

## 1. Introduction

Contemporary spatial planning tries to face the emerging social-environmental crisis by using more sophisticated and participated policies and forms of knowledge, that can be collected through friendly technologies to facilitate interactions.

---

[1] The paper is the result of a common work carried out by the two authors jointly: nonetheless, chapters 1 and 3 have been written by D.Borri, whereas chapter 2 has been written by D.Camarda.
[2] Proofs and reprints should be sent to Domenico Camarda, Dipartimento di Architettura e Urbanistica, Politecnico di Bari, Via Orabona n.4, 70125 – Bari (Italy). Email: d.camarda@poliba.it.

New government arenas and new forums are opened, in which bottom-up adversary movements tend to be replaced by new institutional configurations of governance, including third parties between state and market, variably organized and associated.

Typical interactions, "discourses" of this new type of planning –called either strategic or interactive, communicative, collaborative- call for committed and complex organizations, durable and able to give listening and symbolic answers, more than practical ones. There is the need for learning organizations, new "communities of practice", new understanding of interaction and linguistic mediation, new –also technologic- abilities of recording knowledge and wills, both expert (concentrated) and common (diffused), new abilities of mediation, negotiation, bargain and field agreement.

The important social-environmental objectives of contemporary spatial planning are helped by technological innovation. Noticeable aspects are the growing diffusion of geographical information systems, their increasingly intelligent and interoperable configurations, architectures and functions more and more hybrid, the diffusion of interactive knowledge-acquisition procedures in multi-agent and remote-interaction systems, the return of interest for the procedures of decision and value appraisal-attribution in new and more complex uncertain and multi-agent contexts. In this situation, it is important to develop GISs increasingly dealing with both quantitative and qualitative information, to work upon intelligent systems for monitoring, recognition, classification, decision and management, and to develop organizations and knowledge in complex and often remote interaction networks. Yet, it is important to reflect on the adequacy of technologies to compete with social-environmental problems, that could even be burdened with difficulties to access knowledge and procedures developed in specialized environments, with technologies not sufficiently tested yet. There are some unresolved questions about the collection of knowledge. It is increasingly assumed that "communities of discourse" and "communities of practice" build the space of planning. But the organization of discourse, its rules, opportunities, evolutions, the plurality of discourse agents are still all points to be explored. The same can be said about the communities of practice, whereas today practices in cities are innovated by increasing symbolic transactions and the dematerialization of technologies.

Which tools can be set up to make formal/informal knowledge explicit? The production of new forms of knowledge both at individual levels (single groups, organizations, individuals, communities) and at the level of comprehensive, changing, dynamic organizations still represents a barely known field. Learning-skilled organizations represent the contexts in which processes of production of new knowledge are drawn out, knowledge in use is recorded or external knowledge is acquired.

What is then the role of planners in those production processes: activating or exploring "communities of practice and discourse", mediating or settling down differences, facilitating or surveying processes? Some opportunities are offered to the creation of virtual spaces of interaction by new technologies, that allow the large diffusion of spaces for knowledge interaction, transfer and production. However, there are questions that still remain problematic, such as the ones connected with the forms of representation of knowledge, the methods of exploration, the methods of creating possible syntheses, the forms of interaction.

Is such a deep penetration into communities really effective, in the processes of explicitation, collection and production of knowledge for planning? Virtual spaces still need to be studied with reference to the mechanisms of cognitive interaction activated/activable in multiple environments, with the aim of identifying the transitions among the different levels of the same environment and among different environments -considering such transitions as processes modifying environments themselves and involved relationships, with reference to cognitive and organizational dimensions.

Remarkable problems still remain, such as cognitive scales and hierarchies, and the need to navigate among abstraction levels [20]. They are general problems in intelligent environments but even specific problems in intelligent spatial environments [17], essential for the management of complexity in spatial reasoning [1]. In such virtual spaces it is necessary to support processes of knowledge acquisition, recording, use and transfer with reference to cognitive contexts that are fragmented, conflicting, incomplete, changing and not uniform.

How to represent different types of knowledge in a form that does not alter the mutual mechanisms of relationship which characterize the different forms of interaction? Few agents are more easily referable to coordinate and cooperative behaviours, but they are little representative of complex environments where coordination and cooperation are unstable conditions, that can be achieved through a flow of "critical states".

How to choose the number of possible intelligent agents to be set up in order to support interactions and the preservation/creation of relationships? What are the roles to be assigned to such agents? It is still unclear what kinds of cognitive developments do take place in collaborative multi-agent planning environments, and what are their structures and dynamics.

An important discussion is currently focusing on the real cognitive transformations taking place in multi-agent environments in front of the extreme organizational, procedural and linguistic complexity of such pluri-logic procedures as compared to the traditional mono-logic ones. This literary question appears to be even more remarkable and perhaps more explorable in planning environments, representing organized institutional environments for spatial transformations. In it, the potentials of multi-agency are conditioned by the forms of power distribution, of deliberative practices, of the strong orientation to solution.

In some cases, a useful support for planning is represented by the building up of a shared evaluative and interpretative platform, where the interaction is long lasting, and is commonly defined by scientific literature as a 'double ring'. Within this interaction, it is possible to modify and fine-tune initial positions, in order to allow the convergence toward a unique strategic approach.

In complex regional contexts, and for particularly delicate issues, the 'ring' needs to embrace a number of very skilled and rare knowledge sources, represented by scholars scattered around the world and greatly distant from one another. In such difficult cases, the possibility to reach the desired semantic convergence can be enhanced by using IT-based tools, networks, and dedicated software.

The work presented in this paper is an area-based application of IT-based tools, made up by using MeetingWorks[©], a groupware tool for electronic brainstorming and idea organization, originally designed to facilitating corporate decision making. After this introduction, the first chapter describes the activity of scenario building in the Mediterranean context and its suitability to be aided by the particular IT tool, used for the management of the different forms of knowledge. The chapter also deals with the architecture of the system and the process flow, showing some critical points, both positive and negative, of carrying out a computer-aided participatory session, especially when compared to a traditional face-to-face session. The last chapter draws out some final comments, pointing in particular to some interesting perspectives of IT-based multiple-knowledge management in decision making processes.

**2. IT and environmental policies**

*2.1. Building future scenarios*

Under a project financed by the European Union, oriented to building sustainable development scenarios in the Mediterranean region, some experiences have been carried out since 1999. A Concerted Action of the EU INCO-DC Commission (Dg XII), the project aims at enabling policies for sustainable development in the Mediterranean region, particularly focused on soil and water. The first activity was concentrated on the Tunis case, and its topic was the interplay between agriculture and urbanization; the activity in Izmir dealt with coastal zone management; the third activity in Rabat dealt with globalization vs. local resources, under the prespective of the emerging Euro-Med free trade zone (see the official website http://www.iamb.it/incosusw for details). The scenario-building approach first used in Izmir, Turkey (2001), then being amended and fine-tuned in the case of Rabat, Morocco (2002), is based on a variant to the *strategic choice* approach by Friend and Hickling [7], as developed by the recent evolution of futures studies.

In general, the whole set of future studies methodologies is really multifarious, ranging during its long history from quantitative to qualitative and hybrid methods. Quantitative methods, classically linked to the Delphi iterative method of obtaining shared convergence on issues, do have peculiar military origins. Rigid forecasts and rational predictions were major aims, and their outcomes ended up being particularly appealing also to private entrepreneural management [23].

Because of the increasing awareness of the importance of multiple knowledge in planning processes, together with a larger acquaintance with the underlying presence of uncertainty, then methods tended to evolve towards softer and more hybrid methodologies. Scenario-building approaches begun to combine expert and local-based forms of knowledge, either complementing them with Delphi or moving toward structured but more qualitative knowledge interaction processes [22].

One of the most advanced results has been a pronounced openness to problem-structuring, more than problem-solving approaches. Hence, a higher stress on participatory arenas, both in knowledge sharing and knowledge enhancing interactions, particularly aiming at turning planning and decisionmaking procedures into democratic processes [13].

Participatory arenas have been frequently indicated as essential in the building up of shared future scenarios, with their consensus-enhancing character. In some particular contexts, where problems and expectations are remarkable, scenarios can be gathered through the so-called *future workshops* [12]. The Mediterranean context, with its complexity and current criticality, proved to be particularly suitable to such methodology, and in fact that was the methodology selected by the project steering group.

The *future workshop* methodology is outlined in Table 1. It was selected among other scenario-building methodologies after analyzing the particular implementation context under several points of view. As a result, several reasons were fundamental for the final decision on its utilization, ranging from a strong commitment to action, to a significant involvement of non-experts in designing futures, an enhancement of participants' creativity while building shared visions, a strong orientation to interactive learning, and a useful process flexibility [14].

During the *Critique* phase, stakeholders are asked to describe major structural changes occurred in the area in the past and major problems of the present situation. As Ziegler points out, this involves a basic hypothesis at the very beginning, i.e., that a broad discontent with the current situation does exist, that creates an intense aspiration toward a development change [24]. In fact, contrasting dissatisfactions and desires are supposed to boost the realization of significant future visions. As operational tool, a modified version of Delphi method is applied in order to acquire a common knowledge base on structural changes and current problems for the subsequent phases[3].

In the subsequent *Imagination* phase, the process is carried out in the form of computer-mediated brainstorming session, with the effort to let stakeholders get rid of the current urgencies and to project themselves to the distant future[4]. Traditionally, this phase is carried out as an oral brainstorming session, helped by notes taken throughout the interaction: computer-mediated interactions should then replace informal contacts and notes with a written exchange of ideas. Although consensus on few visions is a desirable goal, computing should not reduce the richness of images and discourses, in order to prevent limitations on future strategies: Delphi method is then at times replaced or hybridized by more argumentative tools [21].

The *Implementation* phase is devoted to the transformation of visions into operational strategies. In this phase, participants are solicited first to reflect on possible actions aimed at implementing their future visions, then to single out policies and related resources in order to enforce those actions. As a result, complex strategies are built up, whose aim is to provide implementation alternatives that should support public decisionmaking for the given time span.

Although not so creative as the fantasy phase, but rather routinary and structurable, this phase is relationally complex and nested and hard to be tackled by computerized interactions. The whole architecture of the process does then show several elements of complexity and hybridation, with particular concentration in the *fantasy* and *implementation* phases.

*2.2. Scenario building in the Mediterranean*

Throughout the INCOSUSW project, future workshops have been set up either as a *vis-à-vis,* computer-free session or as a computer-aided session, with mixed but encouraging results [14]. In the case of scenario building in Izmir, two parallel sessions were developed in the same time: a vis-à-vis session and a computer-aided session. The two sessions were intended as different methods to implement a similar approach in building scenarios, where stakeholders were involved, coming from institutional boards, NGOs and University.

The reasons of exploring the possibility to use a computer program in some phases of the scenario-building activity in Izmir were several. Above all, a major reason was a better manageability of the preliminary Delphi tour for the identification of problematic areas in local coastal zone issues. It was thought that at least some routinary stages, such as the setting up and the exchange of questionnaires among stakeholders, could be accelerated and could facilitate the convergence of outcomes.

After a research on software tools available in the market, useful to assist decisionmaking, to build questionnaires and/or to manage multi-stakeholder meetings, it was decided to choose MeetingWorks©, by *meetingworks.com*. A LAN-based system, MeetingWorks© provides a Chauffeur station for use by the meeting facilitator or an assistant to create a meeting agenda and to run the meeting. Participants have access to Participant stations, where they can enter their ideas, votes, comments and other inputs, all anonymously. The software was conceived as a tool for executive, staff or personnel meetings within firms or societies, where the main aim is to ease the achievement of agreement on projects and issues. In the case of scenario building, instead, the major aim is to enhance the interaction and the share of knowledge from different sources, namely stakeholders, so finally easing the drawing out of alternative community scenarios, development strategies and policies [21]. This difference is important in understanding the pros and cons, as well as the real potential of the tool.

---

[3] In the case of Rabat, the last twenty years were the time span analyzed.

In the last case-study of Rabat, an attempt was made to set up a computer-based interaction throughout the whole process. An initial purpose was to start the interactive session basing on a framework of scenarios previously built up by 'expert' stakeholders. They were to be arranged on the basis of data available from local, regional and national government, as well as from knowledge obtained from scientific and other sources. For each group of experts (Public associations, Scholars and academicians, NGOs), the process leading to scenarios can be summarized and divided into 7 main stages (Figure 1):

1) Reflecting on the concerns raised by an upcoming globalization in local economy, society, environment, as well as on issues and problems generated from that (*Critique*);
2) Generation of desired visions concerning the consequences of globalization, after forgetting the constraints of the present reality (*Fantasy*);
3) Identification of obstacles and limits to the implementation of visions;
4) Definition of possible actions aimed at overcoming obstacles and achieving the desired visions;
5) Definition of resources to be mobilized in order to implement actions;
6) Aggregation of outcomes into two or three alternative scenarios.
7) Grouping homologous scenarios drawn out by groups, in order to have three or four scenarios at the end of the process.

After the completion of the above sequence, scenarios were supposed to be evaluated by a more numerous and assorted group of stakeholders, both experts and non-experts, in order to assess the validity of assumptions, plausibility of development and interpretation of the local context. As a matter of facts, time, organizational and communication constraints among Countries made it impossible to gather the three expert groups: even 'non-expert' (common) stakeholders were difficult to be found in sound number. As a result the three parallel processes were unified and carried out by a single group of assorted stakeholders from institutional, academic and NGO sectors together. With reference to figure 1, the actual process architecture can be referred to the 'Group 1' column only, the last stage being the sixth one.

---

[4] In Rabat, the time horizon of scenarios was the year 2030.

*2.3. The interaction process architecture: limits and potentials*

The Rabat workshop is still under way, so it seems insignificant to discuss deeply the whole experience thoroughly, and to critically compare outcomes with previous case-studies, at this stage[5]. Nonetheless, an overall description of the system architecture for the defining process of scenarios is useful, in order to explore particularly the relational aspects of multi-source knowledge interactions, as supported by computer technology. The description will be carried out with reference to the routine for the first stage (the critique phase, fig.2), under both technical and substantial point of views. The routine pattern is basically similar for each stage, except for some significant adaptations, due to the different themes dealt with. This induces a problem of imperfect repeatability of routines that is addressed with *ad-hoc* approaches, described thereafter.

2.3.1. The basic routine.

The routine flow is made up of 7 consecutive steps, plus some collateral steps (3a to 6a) aiming at monitoring qualitatively some characters of the learning process, both single and collective. The first step is used to write down the concerns over globalization as well as over productive and urban changes as intended by each stakeholder. Lists are prepared singularly and then merged by the software: interactions are anonymous, but participants' responses may be traced back by an ex-post analysis. In Rabat, stakeholders' interaction has been remarkable in this step: it confirms that free and anonymous computer-based interactions allow a more democratic expression of ideas [21], so enhancing responses (63 out of 11 participants). On the other side, this may induce the unmanageability of outcomes, in that it increases the repetition of similar issues, that voice-based interactions might instead filter and prevent.

Therefore, a second step is just needed for the elimination of redundancies, contradictions, inconsistencies among responses, and for the grouping of similar issues under the same heading, in order to have a manageable list for the next steps. In a traditional session this step is typically carried out by means of a general discussion, coordinated by a facilitator and heuristically simplified by informal human interactions. The software does not allow an effective simplification and, therefore, a compromise is currently being carried out, by means of an ex-post simplification of the list, manually grouped and made consistent by the facilitator and his staff. By making this, biases and distortions are not avoided, and the hexogenous synthesis may even create substantial divergences from original ideas. However, this solution seems preferable to the traditional voice discussion because of its better manageability in terms of time and suitability, particularly as regards possible remote, internet-based sessions. In the actual Rabat case, the ex-post grouping was operated by complying with a strict safeguard of the complexity and variety of ideas, so inducing a reduced list of 40 items, with only a 30% reduction rate.

Step 3 is aimed at further reducing the list, but basing on more substantial criteria, able to involve the knowledge-based and experiential values of each single stakeholder. As a matter of facts, they are asked to rate (0-5) the list of issues in terms of importance. Basically, this is the first evaluative step of an iteration, where different rating methods are suggested for each step (steps 4, 5a, 6a, 7), aiming both at singling out the top issues and at controlling the consistency of evaluations. The final purpose is to obtain a manageable list of issues based on a shared consensus, as a result of a mutual learning and awareness-enhancing process, not of a merely reductionist procedure. The produced list is not important *per se*, but in that it involves a learning and knowledge-sharing process, that can be a useful support for the subsequent generation of scenarios [13]. Interestingly, in the Rabat case step 3 did not produce a useful consensus on indicating the most important issues: in fact, three

---

[5] A proper official INCOSUSW report is being prepared on the Rabat workshop, and a further comparative paper will

quarters of all concerns top ranked 4.5 to 3. Admittedly, it is often remarked that pure numbers are not able to properly qualify judgement values, and that such kinds of methodologies are therefore intrinsically erroneous [23]. However, recalling the overall process-oriented aim, this first evaluative step may arguably enhance mutual knowledge exchange and awareness on issues, and this should produce useful effects with subsequent iterations.

A further evaluative step is then step 4, where stakeholders are asked to scroll the list resulted from the preceding step and to choose the 10 most important issues according to their own views, ranking them (1-10). In the Rabat experience, the agreement on a reduced list proved to be difficult, and only 5 out of 40 items where eliminated from the list by the interactive evaluation. As a matter of facts, classical Delphi literature report a formal impossibility of reaching convergence on ideas at an early stage, and the Rabat occurrence confirmed such position [22] [23]. This induces the need of further iterating the evaluation but, in order to overcome associated time constraints, the possibility of disregarding some low ranking issues is envisioned by the process flow.

The dropping-off of the least ranked issues is located in step 5. This phase of selection is manually carried out by the facilitator's staff as a hard cut-off of the top items, basing on the top average rankings. Of course, this operation produces the elimination of ideas that a proper Delphi iteration would instead, probably, enhanced. Being manual and undisclosed, it may jeopardize the original representativeness of issues, especially if least issues remain numerous or there is a high disagreement on them. This last case did occur in the Rabat session, where the 17 selected issues out of 40 were not really founded on a substantial agreement of expressed ideas. This risky operation was then made possible only by the experimental nature of the project, but it can be said that normally such manual simplification may well involve only few low-ranking agreed-upon issues.

In step 6, the iterative process started in step 4 is carried on. In fact, after the first ranking, participants are asked to interact and comment with one another by using an embedded tool of the be forthcoming, following previous ones [2] [14].

computer program: the 'chat'. This unstructured interaction allows a certain relaxation in communication schemes, so emulating a voice discussion that is able to enhance the exchange –not necessarily the production- of ideas. It typically safeguards the democratic expression of ideas, since no precedence is given to fluent catalyzing leaders but the random rapidity in digitizing. Further, the classical criticality implicitly included in any discussion, i.e., the difficulty in taking memory of all issues and the richness of their complex grey shades, is easily overcome by means of the *chat log*, a detailed record of all exchanges. In the Rabat case, the chat tool, intrinsically less structured than other evaluative tools, was enthusiastically accepted by participants, who found themselves more comfortable in exchanging informal comments and ideas. However, outcomes were not significantly effective in enhancing the final consensus in that case, so confirming that when disagreement is high the evaluating cycle should be probably reiterated again and again [15]. In that case a foreseeable risk is casting an excessive time burden over the process flow, and then predetermined time constraints may be useful when using the 'chat' tool, although they may raise doubts of mortifying creativity.

After chatting, if the disagreement on the outcoming list of concerns is low or fairly non-existent, then the last ranked list is the final one of the whole *critique* phase, and the process can move to the next phase of the scenario-building activity. On the contrary, if the desired convergence of outcomes is not reached at the end of step 6, then an iteration is started up in step 7, carried out as *Delphi* tours, in order to reach a possible convergence on a consistent list of items at the very end of the stage. Here a question of time is strongly persistent, and in fact, in the Rabat case, at the end of the secondly iterated step 4, the realization of a further disagreement on the reduction of issues suggested to give up with further attempts and to proceed to subsequent stages with the whole list of 17 resulted issues (concerns).

The above main routine is aimed at structuring the proper scenario-building process throughout all the constituting stages of the future workshop. However, a frequently upcoming critique over the communicative approach behind the process is that not the relevance but rather the argumentative ability of each issue can enhance the achievement of the desired convergence [6]. It is clear that such a critique is able to challenge the whole conceptual framework of building scenarios basing on the mutual exchange and enhancement of single knowledge levels of participants. Should it be possible to have an evidence of actual increases of stakeholders' knowledge levels, then the effectiveness of scenario-building approaches could seem less questionable. In order to try to explore such possibility, some collateral steps are then carried out together with the mainstream of the main routine. In particular, where iterative cycles are set up, the ratings of each stakeholder are taken out after each evaluation step, as ex-post by-products of resulting data (steps 3a and 4a). The longer the iteration, the larger the time series on which to carry out the behavioural analysis. In order to have a more complete set of behavioural data during the whole process, other evaluative steps, less structured than proper rating and ranking steps, are complemented with analyzing tools. In particular, steps 5a and 6a are again aimed at looking into stakeholders' behaviours, even if sympathetically and qualitatively. In these steps, participants are invited to reflect on the impact of the interaction process on their informational level, and to self-rank the level of knowledge gained from the interaction.

However, as a whole, these delicate steps are not yet completely defined at present, and the way how to carry out them effectively is still being explored. As a matter of facts, in the case of Rabat both steps 3a to 6a were not performed at all, because of contingent problems, particularly technical and time.

2.3.2. Notes on *ad-hoc* approaches in subsequent stages.

Basically, the above description reveals the intrinsic impossibility of carrying out a perfectly integrated process, fully automated and properly routinary. When moving from the first stage (*critique phase*) to subsequent phases (*fantasy* and *implementation*), the issues that characterize each phase change substantially, since aims themselves are different from one phase to another. The process is therefore sensibly different from the basic one, with the need of amendments and deviations that induce a substantial hybridization of the whole system architecture.

Far from being useless burdens, such *ad-hoc* approaches aim at increasing the interaction effectiveness, in that they enhance the possibility of letting mutual knowledge exchanges cope with the increased complexity of issues at hand. A typical case is the generation of visions connected with future images of a desired future, dealt with in the *fantasy* phase.

In that phase, creativity needs to be largely enhanced, even to the possible detriment of the optimal manageability of process routines. In fact, experience shows the difficulty of letting people freely express their desires if communication tools act as practical limitation and obstacles of the present reality, so threatening the whole scope of work [14]. In some cases, for example, the imagination phase is carried out by relying almost completely on brainstorming and discussion phases, rather than on Delphi tours. Admittedly, even if computer-aided, discussions are in this case difficult to be recorded properly, and only final agreements and collective outcomes become usable data for subsequent phases, so loosing memory for more detailed information. However, generated images and future visions often prove to be more effective and representative, and more useful to hinge subsequent strategic scenarios on. Particularly complex contexts seem to suggest such a more hybrid approach, unlike communities with more standard and regular everyday problems[6].

In the *implementation* phase, creativity is substantially less crucial than the previous one, and therefore routinary steps may be more adherently performed. However, since the generation of 'obstacles' toward the realization of future visions is followed by the generation of related 'policies' and then 'resources' aimed at overcoming such obstacles, a possible 'combinatorial explosion' may easily make issues unmanageable by the process flow [14]. Possible further simplification and hybridations may then be needed in order to properly manage the process, to the possible detriment of scenarios richness and representativeness. The final outcomes of the Rabat future workshop will soon provide useful feedback in these fields.

*2.4. Significant overall features*

Some positive and negative aspects can be drawn out about the use of such software tools in running computer-aided sessions of scenario-building activity.

First, the use of an ad-hoc computer software was helpful in framing scenarios with robust and credible issues: the anonymous interaction among stakeholders enhanced the free democratic expression of opinions, themes and concerns, particularly important in some Developing Countries.

Secondly, the computer program could ease the real-time setting up of structured scenario alternatives: the main result of accelerated interactions and feedbacks was then to have a resulting data set easy to be read and managed for future policy actions.

A third important feature of a computer-mediated session, when compared to a vis-à-vis session, is that stakeholders can interact without (or at least with a tiny) filter of a human facilitator. This is more likely to happen in a remote interaction, when stakeholders are far from one another, and they are in touch only by means of the computerized session. In this light, MeetingWorks allows the setting up a web-based remote session, either in the same time or in different times, interacting on the basis of an ad-hoc agenda published on an Internet webpage.

A last point is the possibility to check, at least in a qualitative way, what level of knowledge can be raised by the whole iterative process. In fact, the above-mentioned increasing criticism is particularly hard to be challenged, also because in traditional vis-à-vis interactions an evaluation of the level of knowledge, reached either by single participants or by the entire group, is in fact difficult to be verified and likely to be influenced by the interaction itself. As a matter of fact, a computer-based interaction is instead mainly mediated by a software tool, then more likely to minimize biases and mutual influences of stakeholders. Therefore, steps of collective evaluation or self-assessment of the level of knowledge acquired can be more likely to be properly effective.

---

[6] In the Rabat case, an hybridation of voice discussions with chat interactions is being explored.

The above aspects constitute the potential of the use of computer-mediated interactions: however, there are several important critical aspects that need to be addressed, as well.

A first negative point to be underlined is represented by the mediation offered by the computer itself: a cold and non-stimulating medium that is hardly able to make up for warm, informal, sympathetic interactions among humans, as often claimed [3][21]. This occurrence determined several negative by-products, especially after the first intriguing impact: monotony and tedium in replying, scarce involvement, low attention. Shortened stages and discussion intervals can be an answer to such shortcomings, but they are not always applicable in the case of long scenario-building activities, especially if carried out remotely.

A second important negative feature of the software is related to a character that typically automated routines lack, when compared to manual, traditional face-to-face interaction: the control of process flows. In fact, whereas in a traditional interaction among stakeholders issues come out during a continuous process, attended and controlled by the attention of facilitators and stakeholders, in an automated routine processes are often invisible, and there is the risk of hiding steps and generating false heuristics. However, this shortcoming cannot be easily challenged within a software conceived for specific and often undisclosed company processes, such as MeetingWorks. A possible hybridation with human control, or with some other complementary software could be a possible solution that is currently being explored.

A further point is linked to the need of having manageable lists during the process, by eliminating redundancies and grouping similar issues under the same heading. As mentioned before, the software does not allow an effective simplification, and facilitators need to make a selection of issue categories prior to the session (in the Izmir case, these issue areas were five: Political/institutional, Physical/environmental, Social-cultural, Economic and Others -a free field for uncategorizable issues), which is a compromise between a free identification issue areas –leading to a combinatorial explosion- and a stricter classification of ideas. The result is a sequence of several different steps in the process, where all single issues typically need to be individually addressed to each predetermined category. In this case, the high fragmentation of the sequence into micro-steps mutually complementary enhances the complexity of a reduction activity, so involving a long routinary process, able to cause disorientation and loss of attention and interest on stakeholders. Results typically show in the end a high presence of redundant and inconsistent responses, probably stemming from such induced loss of concentration.

In this concern, an automated or semi-automated routine, based on *genetic algorithms*, would be useful, if properly devised [8][11]. In particular, in a first stage, the recognition of issue areas could be derived basing on the analysis of recurrences of concepts by means of GA-funded procedure of text-analysis software tools. In a second stage, the assignments of statements and issues to issue areas could be realized by analyzing the consistency with a representative criteria for each issue. As happening in all classical GA applications, the GA-based routine should then continue until each condition is met, for each statement and toward the completion of each area.

*2.5. Outcomes and remarks*

After some experiences carried with IT-based scenario building, particularly in Developing Countries, some interesting results can be pointed out. First result is that, although not univocally effective in all situations, the computer allowed the carrying out of rapid and remote sessions (especially *Delphi*) hard to be developed alternatively. Secondly, the bunch of resulted datasets is given in real time, which is a remarkable outcome mainly because –above all- it saves time for the next hard decisionmaking and policymaking processes.

Moreover, apart from consolidated results, there are some suggestions that come out particularly in the case-study of Rabat, even if they cannot be claimed as proper results, since the process is not complete. Among a number of intriguing but fuzzy suggestions, one needs to be mentioned in terms of stakeholders' learning behaviour.

In fact, a quick survey of the partial report of the process seems to show that evaluation criteria slightly changed during the various evaluative session. In particular, while mainly financial/economic criteria were used by the group in the very first evaluative step, collective interactions seem to have determined a progressive shift toward social and environmental criteria. Such kind of result, especially if confirmed by the behavioural process of single participants (still under way), can reveal the importance of multi-discipline group interactions in modifying and perhaps socializing the knowledge level of stakeholders participating in the group. Instead of a simple juxtaposition of different languages, peculiar of the cognitive domains that produced them, the interaction may have allowed a certain level of inter-domain transfers, which would be a very interesting perspective in terms of the underlying knowledge-learning process.

Of course, such intriguing results are still vaguely supported at present, and need to be argued and confirmed by further stages of the session and by further analyses and research.

As a whole, the experience carried out seems to suggest new perspectives in the quest for building multi-actor, multi-discipline knowledge bases, oriented to policy scenarios. Especially the obstacle represented by distance seems now to be tackled by web-based computerized interactions.

This occurrence is extremely significant, meaning that in times of globalization, getting multi-discipline knowledge from scattered stakeholders within structured frameworks is not utopian, and its management by local decision makers is not utopian but possible and useful for local communities.

**3. Final comments and research perspectives**

Sustainable planning is currently looking for an equilibrium of interests between the dimensions of "self" and "hethero". Varieties, differences and tones are indispensable ingredients for the coexistence and the development of lives in the society and the environment.

If the "reductio ad unum" induced by globalization is a threat for localisms and biodiversities, globalization involves also the space for new entities: institutions, concepts, information flows, etc. These entities, by introducing new meanings and potentials from different environments, tend to make and keep them crucial for the existence, by making the link between local and global inextricable (an apparent paradox: the search for the "local" derives from the increasing reinforcement of the "global").

"Self" and "hethero" show also more implications in terms of planning, if we look at plans as micro-anticipations and micro-simulations of knowledge-in-practice within evolutionary systems. As planners, we face complex systems, whose values are changing, adaptive and self-organizing. We must structure and solve problems created by those systems. In fact, both in human (individuals, groups, organizations, economies, etc.) and environmental domains, they are dynamic systems involving the cooperation and/or the conflict among multiple agents with multiple cultures.

On the other side, structures and organizations of such problems and systems are created by our dynamic representations and "beliefs". In problem setting and in problem solving a sort of self-organization is involved, due to learning cycles and to the adaptation of the problem through information sharing, conflict mediation and negotiation, reframing.

The role of awareness becomes central. Awareness represents the ability of self-organizing answers, operating through rational/emotional and moral cognition including subjective and qualitative experiences. Awareness means experimenting the "singularity" and thinking of "hethero" as stemming from "self", assuming the raising out of "one from two" and of "two from one". This is a rather mysterious dialectic, even if materialism explains it through the emerging proprieties of neural system.

In the recent mystic vision of problem-making and decision-making, "singularity" is associated with love, and "separateness" with fear. Possibilities of conflict resolution derive from spiritual "singularity" and depend on the ability of different agents to assume their own diversity as different manifestation of "self". Once again, learning processes become crucial, since actors learn their competitors' movements and act consequently.

In such environments there is an equilibrium that is able to self-confirm itself, more than what Nash equilibrium does [18]: i.e., a situation giving also a practical knowledge to individuals/elements facing each other, rather than a mere conceptual and theoretical support to decision.

However, new internet- or intranet-based forms of negotiation still put questions of problem structuring (or pre-structuring), representativeness of actors, comprehension of texts-languages and different cultures, showing the irreducible harshness of every knowledge-in-practice transaction in multi-agent and multi-actor contexts.

Information and communication technologies have impacts on organizational-operational decisions, often giving a decisional environment which is familiar to the more traditional game theory and conflict analysis but requires dynamic approaches careful to non-equilibrium. The development of information and communication technologies and information sciences gives new opportunities to negotiations through forums of "knowledge" and "interests". Negotiations become increasingly multi-cultural and extended in space and time and such expansion creates in turn inter-cultural and intra-cultural contexts in which negotiations take place.

When do practical and successful (fair) negotiations take place in the social-environmental domain planners belong to? How is it possible to adopt integrated negotiation or problem-solving approach (perhaps in a win-win process) to negotiation rather than distributive (win/loose) negotiation? How to build different and compatible objectives? Simple answers do not exist, out of general, somehow moral principles, oriented to putting true cooperation and mutual comprehension at the basis of negotiation.

Current literature outlines that inter-cultural negotiations are difficult and a positive development depends on the knowledge of each party about else's culture, on the way how each negotiator perceives else's behaviour, on implicit objectives emphasized by each culture. The need to avoid the "Babele effect" and the failure (linguistic confusion producing paralysis) caused by generalized incomprehension is still remarkable. During negotiations, cultural differences (global environments) create difficulties not only in understanding words but also in interpreting actions, in giving shape and substance to the given problem and in the selection of the style of negotiation.

According to Faure [5], globalization has brought people closer and supports the management of conflicts through negotiation rather than through the destruction of opponents. As regards negotiation and the role of local/global agents, negotiations have some key elements such as actors, games and strategies (organizations): outcomes (pre- and post-negotiation activities) and in general cultural aspects are crucial.

The model of collective decision by Condorcet is different [4]: "the performance of a group in reaching a correct judgement basing on a majority rule will be superior to individual performances, provided that some conditions are respected". Such conditions need to be explored, and they are generally the following: the probability of a correct decision depends on the technical experience of group voters; voters share a common aim; voters are statistically independent.

Also in Condorcet's model the definition of "correctness" is very complex. In real life it is preferable to replace the concept of correctness with the concept of "expected choice", that is the choice of voters from an infinitely large population within a polling procedure. This is a change that guarantees that decisions are built through a correct process (socially appropriate) and inclusive, which makes use of pertinent data and information. In fact, the process of group consensus represents paths-toward-truth in contexts of incomplete and partial knowledge of discourse-dependent situations.

The integration of multiple-source knowledge as a multiple-objective optimization problem is challenged by large spaces of research and by the difficulty of finding optimal solutions. Moreover, such integration needs to face the continuous risk of redundancy, subsuming and contradiction. It is generally guaranteed by the confidence placed in shaping/solving problems in intelligent environments more than in individual knowledge.

Recent developments of artificial intelligence oriented to integrating knowledge from multiple sources include the use of knowledge systems, cognitive dictionaries, datasets, data dictionaries. In them, every knowledge set is provided with a knowledge dictionary, recording characters and classes of every knowledge share: moreover, even this scheme requires datasets and dictionaries itself. Although intriguing, proposed methods for knowledge integration seem to be less important: for example, genetic algorithms are proposed for the selection of the "optimal" set of rules on the basis of a performance evaluation [9][10]. However, there is no agreement on the integration of different dictionaries regulating the interaction of multiple cognitive sources.

Approaches discussed above operate by using processing technologies of natural language whose aim is to translate the representation of a source into a final representation that can be integrated in other operational systems.

These translations require knowledge structuring through levels of conventional linguistic description.

Today, the use of multiple cognitive sources in planning requires peculiar technological organizations making use, in turn, of tools for processing knowledge avoiding easy validations of models and outcomes. Today's research perspectives in the field of IT-based scenario management and planning are actually oriented to these fronts.

Table 1. The *future workshop* methodology [19].

| Future Workshops | | |
|---|---|---|
| **PHASE** | **CONTENTS** | **EXPECTED RESULTS** |
| **1. Preparation** | The issue to be analysed is decided and the structure and environment of sessions are prepared. | Summary of contributions. |
| **2. Critique** | Clarification, on the issue selected, of dissatisfactions and negative experiences of the present situation. | Problematic areas for the following discussion definition. |
| **3. Fantasy** | Free idea generation (as an answer to the problems) and of desires, dreams, fantasies, opinions concerning the future. The participants are asked to forget practical limitation and obstacles of the present reality. | Indication of a collection of ideas and choice of some solutions and planning guide lines.. |
| **4. Implementation** | Going back to the present reality, to its power structures and to its real limits to analyse the actual feasibility of the previous phase solutions and ideas. Obstacles and limits to the plan implementation identification and definition of possible ways to overcome them. | Creation of strategic lines to be followed in order to fulfil the traced goals. Action plan and implementation proposal drawing. |

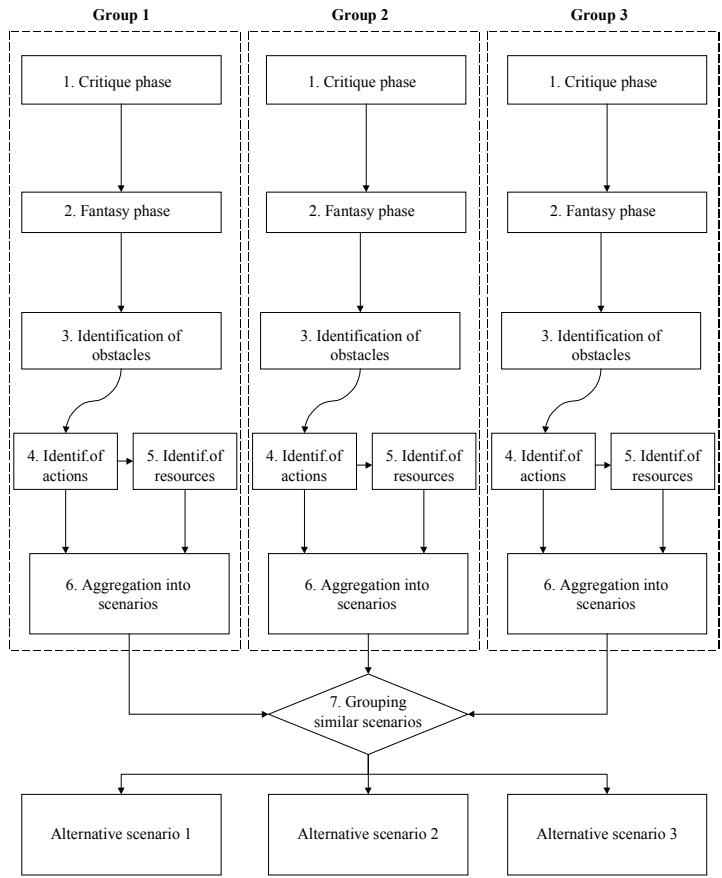

Figure 1. The whole process architecture

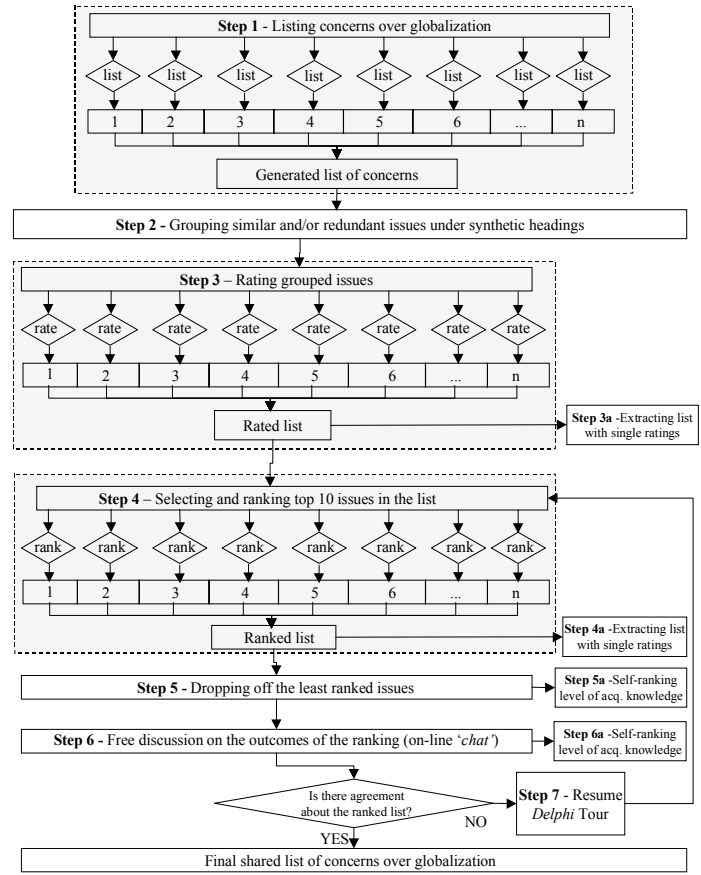

Fig. 2: Critique phase